\documentclass{aa}

\usepackage{graphicx}
\begin {document}

\title{Search for low-mass PMS companions around X-ray selected
late B stars
\thanks{Based on
observations obtained at the European Southern Observatory, La Silla,
Chile (ESO programme No.~62.I-0477, and Swiss 70cm photometric telescope)}}

\author{S. Hubrig\inst{1}
   \and D.  Le Mignant\inst{2}
   \and P.  North\inst{3}
   \and J. Krautter\inst{4}}

\institute{European Southern Observatory, Casilla 19001, Santiago 19, Chile,
\and
 Laboratoire d'Astrophysique, Observatoire de Grenoble, BP53, F-38041 Grenoble, France, \\          
W. M. Keck Observatory, 65-1120 Mamalahoa Highway, Kamuela. HI 96743, USA
\and
Institut d'Astronomie de l'Universit\'e de Lausanne, CH-1290
Chavannes-des-Bois, Switzerland
\and
Landessternwarte Heidelberg-K\"onigstuhl, D-69117 Heidelberg, Germany
}

\offprints{S. Hubrig}

\date{Received  xx /Accepted yy}

\titlerunning{Search for low-mass PMS companions}
\authorrunning{Hubrig et al.}

\maketitle

\abstract{
We have observed 49 X-ray detected bright late B-type dwarfs to search for 
close low-mass pre-main sequence (PMS) companions using the European Southern
Observatory's ADONIS (Adaptive Optics Near Infrared System) instrument. 
We announce the discovery of 21 new companions in 9 binaries, 5 triple, 
4 quadruple system and 1 system consisting of five stars. 
The detected new companions have K magnitudes
between 6$^{\rm m}.$5 and 17$^{\rm m}.$3 and angular separations ranging from 
0.\arcsec2 and 14.\arcsec1 (18-2358 AU).
}

\maketitle
 
\section{Introduction}\label{sec1}

In the ROSAT All-sky Survey (RASS), high X-ray fluxes, up to $10^{31}$
erg/sec, were found for 86 late B-type dwarfs of spectral types B7--B9
(Bergh\"ofer et al. \cite{BSC96}). This was very surprising, since
from a theoretical point of view stars in the spectral range B5-A7
should not have any significant X-ray emission. This was
essentially confirmed for B-type stars by Grillo 
et al. (\cite{GSMVH92}) from their Einstein Observatory SAO-based 
catalog of B-type stars.
In the few cases when X-ray emission had been detected in such B5-A7
stars, it was found to originate from a cool companion.
In order to search for the source of the X-rays, 17 B stars of the RASS
sample which were known to be visual binaries, were observed with
ROSAT's High-Resolution Imager (HRI), which has a spatial resolution
of 5 arcsec. Quite unexpectedly, only in a few cases could the X-ray
emission be linked to the known visual companion (Bergh\"ofer \& Schmitt 
\cite{BS94}, Bergh\"ofer \& Schmitt \cite{BS01}).
Bergh\"ofer et al. (\cite{BSDC97}) noticed that 77\%\ of the RASS detected
stars are known spectroscopic binaries. It is striking that, 
for a large number of these systems, the
observed X-ray fluxes are much too high to be explained by main sequence
SB secondaries of spectral type F, G, K and M 
having L$_X \leq$10$^{29}$ergs$^{-1}$ (Schmitt \cite{Sch97}).
The most likely source of the X-rays might, therefore, be a very active 
late-type companion, such as a PMS star or a very young star on the main
sequence. This hypothesis is also supported by the fact that
a significant fraction of the late-B stars found in the ROSAT
survey belongs to rather young stellar groups: Sco-Cen association, Sco
OB2, and the Pleiades cluster and supercluster.

It was shown that approximately two-thirds of all solar-type field
stars are members of multiple systems (e.g. Duquennoy \& Mayor
\cite{DM91}), and the fraction of stars which are formed in binary or
multiple systems may be even higher (e.g. Ghez et al. \cite{GNM93},
Leinert et al. \cite{LZW93}).  However, only a few systems with highly
differing masses are known.  A pioneering study had been carried out
by Gahm et al. (\cite{GAL83}) and by Lindroos (\cite{L86}) who
found 78 likely pairs with an early-type and a late-type
companion. X-ray observations from a part of the Lindroos systems were
carried out by Huelamo et al. (\cite{HNSSZ00}); they found several of
the late-type companions to be {\sl bona fide} PMS stars. 

The scientific interest in binarity studies of higher mass stars
derives from the question of whether the different processes of high-mass 
and low-mass star formation (Shu et al. \cite{SAL87}) reflect 
themselves in different binary frequencies and distribution of mass ratios.

One method to study whether late B-type stars are intrinsic X-ray emitters or
whether the detected X-ray emission originates from a hitherto undetected
companion is diffraction-limited imaging in the optical
or infrared spectral range on a sample of 
X-ray selected late B-type stars. 
Optical speckle methods are not suitable for the detection of possible
X-ray emitting PMS companions, as the magnitude differences
between the bright early-type primaries and the faint late-type secondaries
in the visual is rather extreme, i.e. of the order of 8-10 magnitudes.
However, in the near-infrared (J,H,K) the contrast between the primary and
secondary components is much more favorable, i.e. only 4-5 magnitudes 
difference.
This much improved contrast in the infrared and the fact that the primaries
are very bright visual objects (V=1$^{\rm m}$.8--6$^{\rm m}$.5) 
made this a perfect project for
high angular resolution imaging using a high order adaptive optics system.

In Sect. 2, we describe the observations performed with the ADONIS instrument  
and the data reduction procedure. In Sect. 3, we present the results.
In Sect. 4, we discuss the implication of these results for the 
formation and evolution of binary and multiple stars.

\section{Observations and data reduction}\label{sec2}

\begin{table}
\caption{The list of X-ray selected late B-type stars}
\label{tab1}
\begin{tabular}{rclcl}
\hline \\[-4pt]
\multicolumn{1}{c}{HD}&\multicolumn{1}{c}{$V$}
&Sp. type&\multicolumn{1}{c}{$\log$ L$_X$$^{1}$}&\multicolumn{1}{c}{Remarks}\\[4pt]
\hline \\[-4pt]
1685&5.50&B9V&29.89&CM$^{\mathrm{a}}$\\
21790&4.73&B9V&29.47&CM\\
27376&3.55&B9Vmnp&28.54&CM, SB2$^{\mathrm{b}}$ + vis.comp.$^{\mathrm{c}}$\\
27657&5.87&B9IV&29.76&CM, vis.comp. \\
29589&5.45&B8mnp&30.74&CM, SB1 $^{\mathrm{d}}$\\
32964&5.10&B9mnp&29.80&SB2 + vis.comp. \\
33802&4.46&B8V&30.75&vis. comp.\\
33904&3.28&B9mnp&30.02&CM\\
36408&5.46&B7III&29.64&SB1 + vis.comp. \\
39780&6.18&B9.5IV&29.44&CM, SB2 + vis.comp.\\
40964&6.59&B8V&30.23&vis.comp.\\
49333&6.08&B7Hew&30.31&\\
50860&6.47&B8V&30.71&vis.comp.\\
73340&5.78&B8p&30.38 &\\
73952&6.45&B8V&29.73&vis.comp.\\ 
74067&5.20&B9V&29.27&vis.comp. \\
75333&5.31&B9mnp&30.04&\\
78556&5.61&B9.5Si&29.72&SB1 + vis.comp.\\
79469&3.88&B9.5V&29.20&SB1 + vis.comp. \\
83944&4.51&B9IV-V&29.29&SB2 \\
90972&5.57&B9.5V&29.53&SB2 + vis.comp.\\
92664&5.50&B9Si&29.78&\\
100841&3.12&B9III&29.38&vis.comp.\\
108767&2.95&B9.5V&28.71&vis.comp.\\
110073&4.63&B8mnp&30.13&SB1 \\
113902&5.71&B8V&30.42&\\ 
114911&4.77&B8V&30.10&SB2 + vis.comp.\\
118354&5.89&B8V&30.10&\\
119055&5.75&AIV&29.94&vis.comp.\\
126981&5.50&B8V&30.07&\\
129956&5.67&B9.5V&29.28&\\
133880&5.76&B8IV&30.29&\\
134837&6.08&B8V&29.70&\\
134946&6.47&B8III&30.00&vis.comp.\\
135734&4.27&B8V&29.97&vis.comp. \\
141556&3.95&B9hgmn&29.79&SB2 \\
144334&5.92&B9Si&30.19&\\
145483&5.67&B9V&30.41&vis.comp.\\
158094&3.60&B8V&29.04&vis.comp.\\
159376&6.48&B9Si&30.14&\\
169022&1.80&B9.5III&29.99&vis.comp.\\
169978&4.61&B7.5III&29.99 &\\
176270&6.40&B8IV-V&30.43&SB1 + vis.comp.\\
177756&3.43&B9V&28.47&\\
180555&5.63&B9.5&29.45&vis.comp.\\
181869&3.95&B8V&29.57&SB1 \\
184707&4.60&B8/B9V&30.40&vis.comp.\\
221507&4.37&B9.5mnp&29.58&CM\\
222847&5.24&B9V&29.78&CM\\

\hline \\[-4pt]
\end{tabular}
$^{1}$ $\log$ L$_X$ values are taken from Bergh\"ofer et al. (\cite
{BSC96})
\begin{list}{}{}
\item[$^{\mathrm{a}}$] CM  : coronographic mode
\item[$^{\mathrm{b}}$] SB2 : double-lined spectroscopic binary
\item[$^{\mathrm{c}}$] vis.comp. : the visual companion already known from the
literature
\item[$^{\mathrm{d}}$] SB1 : spectroscopic binary
known from the literature
\end{list}
\end{table}

A total of 49 late-type X-ray emitting B dwarfs 
were observed. The observations were performed
with the ESO adaptive optics system ADONIS and SHARPII+ (System for
High Angular Resolution Pictures) near infrared (NIR) imaging camera
with a NICMOS III 256$\times$ 256 pixel array at the 3.6-m telescope
with a plate scale of 0.050 arcsec/pixel, 
yielding a camera field-of-view of 12.\arcsec8$\times$12.\arcsec8.
Since all observed objects are bright 
stars they were used as reference stars for wavefront sensing. 
Forty stars were observed in the normal imaging mode in the K band during two 
nights in March 1999.
Whenever the star appeared to be multiple it
was immediately observed in the J and H bands. 
The K-band survey was the priority during these 
nights. A few companions could be  detected from K band images 
only during the later careful data reduction. For those objects only K-band
images are available.

The AO system and NIR camera settings time overheads are of the order of a few 
minutes for every target in such a survey. Given the time constraints, we 
decided: 1) not to optimize the AO system 
performance when switching to the next target and the result of such a
choice was the lower
resolution in J and H bands; 2) to spend minimum time to adjust 
individual integration time, although the PSF peak intensity may vary by more 
than 
10\% between consecutive frames.
Therefore, for a few systems the central pixels in the image 
core of the primary star are saturated. As a result,
the photometric calibration for these stars provides only a lower
limit of the K magnitude.\\
Every target was observed at five positions on the NIR array using a dither 
pattern. As a result, the size for the observed area is of 
24\arcsec$\times$24\arcsec\ centered on the target.
For every position we recorded
16 frames with an individual integration time between 300 and 1200 ms, 
depending on the star brightness. For the stars with detected companions,
we have subsequently observed calibrator stars for 
point spread function (PSF) measurements and 
{\it a posteriori} deconvolution. Photometric standards were observed under the
same conditions to get absolute JHK photometry.
As measured by the La Silla Differential Image Motion 
Monitor (DIMM), the atmospheric seeing during our run was varying with time, 
ranging from 0.\arcsec8 to 1.\arcsec2.
In the K-band, the diffraction limit was reached at all times 
(FWHM $\approx$ 0.\arcsec13), with the Strehl ratio ranging from 15 to 
35\%. 
In the J and H bands the images are not diffraction-limited with
the angular resolution varying from 0.\arcsec13 to 0.\arcsec20. Additionally, 
9 late-type B X-ray emitting B dwarf stars were imaged with the coronographic 
mask during the night of 26 September 1999. Thus, the whole sample of 
studied stars consist of 49 objects. 

In Table 1 we present the list of the observed
late B-type stars. Their visual magnitudes and spectral types
were retrieved from the SIMBAD data base.  The X-ray luminosity $\log$
L$_X$ given in column 4 was taken from Bergh\"ofer et al. (\cite
{BSC96}). In the last column we give some remarks about the observing mode
and binarity.

For the data reduction we used our own C-shell scripts based on the 
ESO/{\bf eclipse} data reduction C routines (Devillard~\cite{DE97}). 
The background emission was computed for every pixel as a median value of 
the stack of 16$\times$5 frames. For each exposure we subtracted background sky
emission, divided by flat field, corrected for bad pixels and used 
shift-and-add method to yield the final images.

The photometry was performed using the IRAF/DIGIPHOT package. For the systems
with an angular separation larger than 2\arcsec\ we applied a 
standard aperture photometry method.
For other systems we combined PSF fitting of the components and 
aperture photometry to get the absolute fluxes of the stars. 

For a few systems  with angular separation smaller than 0.\arcsec4, we 
cross-checked PSF fitting with PLucy deconvolution, which is provided by
IRAF Space Telescope Science Data Analysis System. 
No discrepancy was found between PSF fitting 
and PLucy deconvolution results, either for photometric fluxes or for star 
astrometric positions.
The average rms error is 0.03\, mag for differential photometry, 0.05\, mag
for absolute photometry, 0.\arcsec005 on the projected separation, and  
0.2$^\circ$
on the position angle.
For objects observed in the coronographic mode,
the average rms error for absolute photometry is 0.1 mag,
0.\arcsec05 for the projected separation and 0.2$^\circ$ for the position 
angle, respectively.

The detection limits for these observations depend 
on the separation range, as we cannot detect 
any companions below 0.1\arcsec, and on the contrast between primary and 
companion(s). This contrast also depends on separation:
photon noise prevents from detecting very close binaries (0.1 to 0.3\arcsec), 
while detector read-out noise is the limiting factor for wider systems.
The detection limit values estimated for our observational data set are: 
$\Delta$K $\sim 2$ at separations of $\sim 0.3$\arcsec\ and below, 
$\Delta$K $\sim 5$ at separations $\sim 2$\arcsec and $\Delta$K $\sim 9$ at
separations of $\sim 4$\arcsec\ and beyond.

Most stars had been measured in the GENEVA photometric system (Golay \cite{G80};
Rufener \& Nicolet \cite{RN88}; Cramer \cite{C99}), with the photoelectric
photometer P7 (Burnet \& Rufener \cite{BR79}) installed on the 70cm Swiss
telescope at La Silla (ESO, Chile). The photometric data in the GENEVA system
are collected in the General Catalogue (Rufener \cite{R88}) and its up-to-date
database (Burki et al. \cite{B01}).

\section {Results}\label{sec3}

For 25 of our 49 X-ray selected late B-type stars we could find companions
in the scanned 24\arcsec x 24\arcsec area. Six of these systems were already 
known as visual binaries.
The discovered companions have K magnitudes between 6$^{\rm m}.$5
and 17$^{\rm m}.$3 and angular separations ranging from 0.\arcsec2 to
14.\arcsec1.  

{\footnotesize

\begin{table*}
\caption{ Photometry and astrometry of X-ray selected late B-type stars with
companions}
\label{tab2}
\begin{tabular}{|c|c|c|c|c|c|c|c|c|c|}
\hline
\multicolumn{10}{|c|}{Detected visual systems}\\
\hline
HD &  Comp. & $\rho$('') & P.A. ($^\circ$) & J & H & K & 
$\Delta$ K & J-H& H-K\\
\hline
\hline
HD 1685 & A & &&&&&&&\\
       & B & 2.28 & 211.4 & 11.7 & 10.3 & 10.1 && 1.4 & 0.2\\
\hline
 HD 21790 & A &&&&&&&&\\
          & B & 14.12 & 114.5 &&&16.2 &&& \\
\hline
HD 27376 & A & &&&&&&&\\
         & B & 5.32 & 162.5 & 10.7 & 10.0 & 9.9 && 0.7 & 0.1\\
\hline
HD 29589 & A & &&&&&&&\\
         & B &10.00 &251.3 &&& 17.3 &&&\\

\hline
HD 32964 & A & & & $\leq$ 5.19 & $\leq$ 5.22 & $\leq$ 5.22 &&&\\
         & B & 1.613 & 232.6 & 10.03 & 9.37 & 9.38 &$\geq 4.16$& 0.66 & 0.01
\\
\hline
HD 33802 & A & & & 4.73 & 4.78 & 4.78 && -0.05 & 0.0 \\           
         & B & 0.371 & 22.9 & 6.92 & 6.90 & 7.22 &2.44& 0.02 & -0.32\\
\hline
HD 73340 & A & && 5.97 & 6.10 & 6.13 && -0.13 & -0.03 \\
         & B & 0.604 & 221.2 & 9.32 & 8.71 & 8.65 &2.52& 0.61 & 0.06\\
\hline
HD 73952 & A &&& 6.62 & 6.57 & 6.63 && 0.05 & -0.06 \\
         & B & 1.162 & 205.3 & 11.8 & 11.25 & 10.82 &4.19& 0.55 & 0.43\\
\hline
HD 75333 & A &&& 5.40 & 5.45 & 5.49 && -0.05 & -0.04\\
         & B & 1.340 & 165.8 & 10.21 & 9.42 & 9.45 &3.96& 0.79 & -0.03\\
\hline
HD 110073 & A & && 4.74 & 4.78 & 4.85 && -0.04 & -0.07\\
          & B & 1.192 & 75.0 & 8.29 & 8.00 & 7.93 &3.08& 0.29 & 0.07\\
\hline
HD 114911 & A &&& 5.06 & 5.07 & 5.11 && -0.01 & -0.04\\
          & B & 2.706 & 124.6 & 10.31 & 9.61 & 9.43 &0.319& 0.70 & 0.18\\
\hline
HD 133880 & A &&& 5.94 & 5.99 & 6.06 && -0.05 & -0.07\\
          & B & 1.222 & 109.2 & 9.01 & 8.55 & 8.41 &2.35& 0.46 & 0.14\\
\hline
HD 134837 & A &&&&& $\leq$ 6.33 & &&\\
          & B & 4.696 & 154.3 & & & 10.99 &$\geq$4.66& &\\                      
\hline
HD 134946 & A &&&&& $\leq$6.47 & &&\\
          & B & 8.212 & 45.3 & & & 12.51&$\geq$6.04 &&\\  
\hline
HD 135734 & A & &&&& 5.11 & &&\\
           & B & 1.038 & 322.5  & && 5.32 &0.21&&\\
           & C & 6.146 & 156.9 &&& 14.7&9.59&&\\
\hline
HD 145483 & A & & &&& $\leq$ 5.94  &&& \\
                 &  B & 3.767 & 70.8  & && 6.85 &$\geq$ 0.91&&\\
                 &  C & 0.202 & 39.16 &&& 7.94 &$\geq$2.00&&\\
\hline
HD 159376 & A & &&&& $\leq$ 6.32 & &&\\
                & B & 8.703 & 52.9 &&& 12.66 &$\geq$6.34&&\\
                & C & 5.420 & 10.4 &&& 14.18 &$\geq$7.86&&\\
                & D & 6.340 & 141.6 &&& 15.07 &$\geq$8.75&& \\ 
\hline
HD 169022 & A & && 1.79 & 1.80 & 1.82 && -0.01 & -0.02 \\
                & B & 2.392 & 142.3 & 6.67 & 6.45 & 6.50 &4.68& 0.22 & -0.05\\
\hline
HD 169978 & A & &&&& $\leq$ 4.92 & &&\\
                &B & 3.085 & 168.7 &&& 13.69 &$\geq$8.77&&\\
\hline
\multicolumn{10}{|c|}{Stars already known to be members of visual binary systems}\\
\hline
HD 27657 & A & & & && 7.0 & &&\\
         &B & 4.131 & 2.8 & && 8.15 &1.15&&\\
\hline   
HD 36408 & A & &&&& 6.09 & &&\\
            & B & 9.742 & 141.0 &&& 6.34 &0.25&&\\
\hline
HD 78556 & A &  && && 5.65 & && \\
           & B & 1.300 & 298.5 & && 8.96 &3.31&&\\
\hline
HD 90972 & A & & & && 5.63 & &&\\
               & B & 11.084 & 225.8 & && 8.34 &2.71&&\\
\hline
HD 119055 & A & & & && $\leq$ 5.73 & &&\\
               & B & 4.688 & 134.2 & & &8.21 &$\geq$2.48&&\\
\hline
HD 184707 & A & &&&& $\leq$ 4.83 &&&\\
                & B & 2.435 &173.1&&& 7.59 &$\geq$2.76&&\\
\hline
\end{tabular}         
\end{table*}}

The photometric and astrometric results of
our observations are presented in Table 2, where we list the HD numbers; the
separations and position angles; J, H and K magnitudes; the magnitude
difference $\Delta$K between the primary and the companion in the
K-band; and the $J-H$ and $H-K$ colors.
Only in two systems -- \object{HD 1685} and \object{HD 73952} -- do the
companions have a considerable [$\Delta(H-K)\geq 0.2$~mag] NIR excess. Usually,
NIR excess betrays the existence of substantial amounts of circumstellar
matter around the stars.

\subsection{Probability for random pairing}
In order to check whether some of our late-type binary identifications
could be due to a random positional coincidence with field stars, we
calculated the probability P that a stellar K-band source occurs by random
coincidence in the total area of our image fields at given K magnitude.

For our calculations we used the density of infrared K-band sources as
measured by both DENIS (Deep Near-Infrared Survey of the Southern Sky;
Epchtein~\cite{E98}) and 2MASS surveys (Two Microns All Sky
Survey\footnote{see http://irsa.ipac.caltech.edu/applications/Stats};
Skrutskie~\cite{S2000}). For DENIS, it was not possible to calculate
individual numbers for the different galactic latitudes of our target
stars, since star counts from this survey have so far been published only for a
few selected fields. In order to get conservative upper limits,
we used the field with the highest K-band star count density 
we found in the literature. This is a one-square-degree field centered
around the galactic coordinates l=331$^{o}$, b=-1.81$^{o}$ (Ruphy
\cite{R98}).  This field is in a very crowded area close to the
galactic plane.

The probabilities for the presence of a background star for our total
ADONIS image field of view of 24\arcsec $\times$ 24\arcsec\ is presented
in Table 3 for a number of K-band magnitudes.

For 2MASS, we could compute the probability of finding a companion for 14
out of the 19 systems found, by counting stars of a given magnitude in a circle
centered on the primary and with a $1\degr$ radius. Sorting the companions by
their $K$ magnitudes and selecting the largest probability in each
$0.5$ magnitude bin, we obtain the numbers listed in Column 3 of Table 3.
The very large probability given for $K=13$ comes from \object{HD 159376}, which
lies in an exceptionally crowded field.
\begin{table}[ht]
\caption[]{Upper limits for probabilities P of presence of background 
sources for various K-band magnitudes.}
\label{tab3}
\begin{tabular}{rll} 
\noalign{\smallskip} \hline \noalign{\smallskip}
K [mag] & P(DENIS) & P$_{\mathrm max}$(2MASS) \\ 
\noalign{\smallskip} \hline \noalign{\smallskip} 
 6  &  0.0004 & 0.0007 \\
 7  &  0.002 & 0.0002 \\
 8  &  0.006 & 0.0007 \\
 9  &  0.012 & 0.001  \\
10  &  0.028 & 0.001  \\
11  &  0.067 & 0.009 \\
12  &  0.116 & --  \\
13  &  0.200 & 0.93 \\
\noalign{\smallskip} \hline
\end{tabular}
\label{idents}
\end{table}

The DENIS numbers are very conservative upper limits only, since they
were calculated on the basis of most pessimistic assumptions.
First, all of our 
objects are located at
higher galactic latitudes where the star counts drastically
decrease. For instance, as shown by Ruphy et al. (\cite{REC97})
for areas located at $l=303\degr$ the star counts decrease by about
a factor of 10 between $b=2.37\degr$ and $b=-14.63\degr$. Second, they
were calculated for the whole field of 576 arcsec$^{2}$, whereas
the mean distance of the companion stars found is 4.3 arcsec only.

It is worth noting that out of 29 detected companions in our 
24\arcsec$\times$24\arcsec\ field K images, 
17 are found at a distance less than 5\arcsec\ from the primary
(note that 8 out of the 29 companions detected were already known).
And 27 of the 29 detected companions are at distances smaller than or equal
to 10\arcsec. In other words, 93\% of the detected companions are 
found in the 55\% central part of the scanned area around the primary.
Our estimate is that on average the
probabilities for a random positional coincidence of our sources 
are at least a factor of two lower than the numbers in Table 3.
In view of these facts, we can safely assume that -- with the exception
of the six sources fainter than 13th magnitude -- random positional
coincidences do not play a significant role. For 
\object{HD 21790} and \object{HD 29589} where objects with K=16.2 and K=17.3
were found at relatively large distances of 14.\arcsec12 and 10\arcsec\, arcsec, 
respectively, a random positional coincidence can definitely not be excluded.
These conclusions are entirely confirmed by the 2MASS results, which
add much weight to them.

\subsection{Fundamental parameters of primaries}
In Table 4 we present the basic data for the late-B primaries.
We  give the distances $d$ calculated from Hipparcos parallaxes, 
relative uncertainties of parallaxes, 
absolute visual magnitudes, bolometric luminosities,
effective temperatures, masses (in solar masses) and the ages. 
We only accepted those parallaxes which exceed their corresponding error by 
at last a factor of three. This criterion is not fulfilled only for one star
in our sample, \object{HD 36408}, which was already known to have a visual
companion at the distance 9.\arcsec74  before our observations. In the last
column  we give values for L$_X$. We use the standard equations 
(see e.g. Bergh\"ofer et al. \cite{BSC96}, equations 2 through 5)
to calculate L$_X$, M$_V$ and ${\rm L}_{\rm Bol}$.
For the calculation of L$_X$ we use Hipparcos values for the distances and 
$f_X$ values from Bergh\"ofer et al. (\cite{BSC96}).
We assume R=3.1 for the standard interstellar reddening law in which 
the $E(B-V)$ values have been calculated from $(B-V)$ colors taken
from the Bright Star Catalogue (\cite{HJ82}, henceforth {\bf  BSC}) and from
intrinsic $(B-V)_\odot$ colors for late B-type
stars taken from FitzGerald (\cite{F70}).
Finally, the bolometric luminosities were calculated using bolometric
corrections published by Schmidt-Kaler (\cite{SK82})

{\footnotesize

\begin{table*}
\caption{Fundamental parameters of late-B primaries with good Hipparcos
parallaxes}
\label{tab4}
\begin{tabular}{|c|c|c|c|c|c|c|c|c|}
\hline
\multicolumn{9}{|c|}{Detected visual systems}\\
\hline
HD & $d$ (pc) & $\sigma(\pi)/\pi$ & M$_V$ & $\log(L/L_\odot)$&
 $\log(T_{\rm eff})$& M/M$_\odot$  & $\log$ age  & $\log$ L$_X$$^{1}$ \\
\hline
\hline
HD 1685 & 93.9 & 0.048 & 0.57 & $1.870\pm0.068$ & $4.016\pm0.006$ &$2.717\pm0.080$&$8.433\pm0.032$ & 29.79 \\
\hline
HD 21790 & 116.7 & 0.085 & -0.68 &$ 2.473\pm0.088$&$4.064\pm0.002$ &$ 3.659\pm0.147$ &
$8.251\pm0.023$ & 29.83\\
\hline
HD 27376 & 54.7 & 0.030 & -0.20 &$2.378\pm0.049$ & $4.106\pm0.002$&$3.640\pm0.069$ &
$8.107\pm0.022$ & 28.71 \\
\hline
HD 29589 & 105.7 & 0.082 & 0.19 &$2.344\pm0.085$ &$4.161\pm0.003$&$3.876\pm0.127$ &
$7.041\pm1.158$ & 30.33 \\
\hline
HD 32964 & 85.8 & 0.063 & 0.37 &$2.012\pm0.073$ &$4.045\pm0.002$&$2.953\pm0.084$ 
&$8.319\pm0.032$ & 29.76\\     
\hline
HD 33802 & 73.9& 0.051 & 0.01 &$2.309\pm0.060$ &$4.117\pm0.002$&$3.602\pm0.078$ &
$7.974\pm0.067$ & 30.68 \\           
\hline
HD 73340 & 143.1 & 0.063 & -0.13 &$2.511\pm0.072$ &$4.178\pm0.003$&$4.226\pm0.111$ &
$7.454\pm0.274$ & 30.34 \\
\hline
HD 73952 &154.8 & 0.076 & 0.40 &$2.097\pm0.072$&$4.087\pm0.002$ &$3.207\pm0.083$ &
$8.030\pm0.111$ & 29.50 \\
\hline
HD 75333 & 134.2 & 0.099 & -0.46 &$2.441\pm0.099$&$4.088\pm0.002$ &$3.670\pm0.155$ &
$8.192\pm0.008$ & 30.29 \\
\hline
HD 110073 & 108.8 & 0.092 & -0.71 &$2.585\pm0.087$&$4.111\pm0.002$ &$3.975\pm0.151$ &
$8.110\pm0.010$ & 29.57\\
\hline
HD 114911 & 124.4 & 0.073& -0.81 &$2.622\pm0.075$&$4.106\pm0.002$ &$4.026\pm0.136$ &
$8.118\pm0.012$ & 29.95\\
\hline
HD 133880 & 126.6 & 0.108 & 0.20 &$2.195\pm0.134$&$4.079\pm0.011$ &$3.290\pm0.190$ &
$8.192\pm0.071$ & 29.94 \\
\hline
HD 134837 & 111.1 & 0.092& 0.74 &$1.974\pm0.084$&$4.097\pm0.002$ &$3.111\pm0.101$&
$7.022\pm2.126$ & 29.35\\
\hline
HD 134946 & 126.4& 0.119& 0.55 &$2.194\pm0.108$&$4.139\pm0.002$ &$3.564\pm0.147$&
$6.152\pm4.000$ & 29.79 \\
\hline
HD 135734 & 89.1 & 0.086 & -0.57 &$2.517\pm0.086$&$4.105\pm0.002$ &$3.843\pm0.139$&
$8.134\pm0.008$ & 30.12\\
\hline
HD 145483 & 91.4 & 0.083 & 0.56 & $1.960\pm0.084$&$4.057\pm0.003$ & $2.939\pm0.088$&
$8.177\pm0.103$ & 30.30 \\
\hline
HD 159376 & 271.0 & 0.217 & -1.24 & $2.732\pm0.210$&$4.072\pm0.011$ & $4.099\pm0.420$&
$8.178\pm0.078$ & 30.49 \\
\hline
HD 169022 & 44.3 & 0.045 & -1.41 & $2.560\pm0.070$&$3.960\pm0.004$ &$3.515\pm0.138$ &
$8.366\pm0.045$ & 29.34 \\
\hline
HD 169978 & 146.8 & 0.110 & -1.31 & $2.819\pm0.102$&$4.106\pm0.002$& $4.392\pm0.216$ &
$8.080\pm0.034$ & 29.98\\
\hline
\hline
\multicolumn{8}{|c|}{Stars already known to be members of visual binary 
systems}\\
\hline
HD 27657 & 142.0 & 0.082 & -0.04 & $2.253\pm0.086$&$4.078\pm0.002$ & $3.363\pm0.117$&
$8.219\pm0.023$ & 29.59\\
\hline   
HD 36408 & 342.5 & 0.538 &-2.60 &$2.705\pm0.165$&$4.087\pm0.002$&$4.109\pm0.332$&$8.150\pm0.054$&30.05 \\
\hline
HD 78556 & 275.5 & 0.267 & -1.65 & $2.727\pm0.275$&$4.000\pm0.013$ &$3.853\pm0.604$ &
$8.260\pm0.177$ & 30.29 \\
\hline
HD 90972 & 147.5 & 0.109 & -0.37 & $2.303\pm0.101$&$4.046\pm0.003$ &$3.350\pm0.153$ &
$8.330\pm0.020$ & 29.86\\
\hline
HD 119055 & 92.8 & 0.084 & 0.79 & $1.745\pm0.088$&$3.994\pm0.004$ & $2.536\pm0.093$&
$8.509\pm0.036$ & 29.94\\
\hline
HD 184707 & 58.0 & 0.052 & 0.75 & $1.859\pm0.079$&$4.045\pm0.006$ &$2.794\pm0.090$ &
$8.181\pm0.114$& 30.22\\
\hline                                                                       
\end{tabular} \\

$^{1}$ $\log$ L$_X$ values are recomputed using Hipparcos values for the 
distances and $f_X$ values from Bergh\"ofer et al. (\cite{BSC96})

\end{table*}}

The effective temperatures of all stars but one (\object{HD 119055}) were  determined
using Geneva photometry calibrated by K\"unzli et~al. (\cite{KNKN97}), assuming
a metallicity [M/H]=0.0. The most appropriate grid was chosen, i.e. that of the
reddening-free $X$ and $Y$ parameters, which are mostly sensitive to
$T_{\rm eff}$ and luminosity respectively (Cramer \& Maeder \cite{CM79}). The
formal errors given by the code are propagated from typical photometric errors
and are generally lower than 100~K. They are optimistic in the sense that they
are only random errors and do not include possible systematic errors linked
with the calibration or with cosmic causes such as unresolved duplicity.
In the case of the Bp Si stars \object{HD 78556}, \object{HD 133880} and
\object{HD 159376}, the
temperature was determined using the recipe given by North (\cite{N98}), still
using Geneva photometry. For them, the error on $T_{\rm eff}$ was
arbitrarily set to 300~K because the temperature estimate is less reliable than
for normal stars.

The effective temperature of \object{HD 119055} was determined by interpolating
in the uvby$\beta$ grid of Str\"omgren indices computed for [M/H]=0.0,
from ATLAS9 models and fluxes (Castelli \cite{C00}).
Observed Str\"omgren indices were taken from the 
catalogue of Mermilliod et al. (\cite{MMH97}) and were dereddened using 
the UVBYLIST code of Moon (\cite{M85}).
The errors associated with the parameters were derived
by assuming an uncertainty of $\pm 0.015$ mag for all the observed indices,
except $\beta$, for which a probable error of $\pm$ 0.005 mag was adopted.
The accuracy of the determination of the effective temperatures based on
Str\"omgren photometry is better than 100~K.

For the star \object{HD 36408}, which has a very bad parallax accuracy, all 
fundamental parameters are based on purely photometric estimates.

To derive the masses and ages of late B-type primaries, we used the grids
of stellar models provided by Schaller et al. (\cite{SSMM92}). The luminosities
were estimated from the Hipparcos parallaxes (except for \object{HD 36408}, for which
they were obtained from photometric $T_{\rm eff}$, $\log g$ and evolutionary
tracks) and from
visual absorption obtained from Geneva $X$ and $Y$ photometric parameters
through the calibration of Cramer (\cite{C82}), assuming $A_v/E[B-V]=3.65$.
The bolometric corrections are those of Schmidt-Kaler (\cite{SK82}).

The masses were interpolated in the evolutionary tracks (using a code by PN)
while the ages were interpolated in the isochrones of the same models (using a
code kindly made available by Drs. C. Jordi and F. Figueras).
Fig. 1 shows the positions of the late B-type primaries in the H-R diagram.
For comparison we show also the evolutionary tracks of stars with masses 
of 2.5, 3.0, 4.0 and 5.0 M/M$_\odot$.

\begin{figure}
\resizebox{\hsize}{!}{\includegraphics[width=9.0cm]{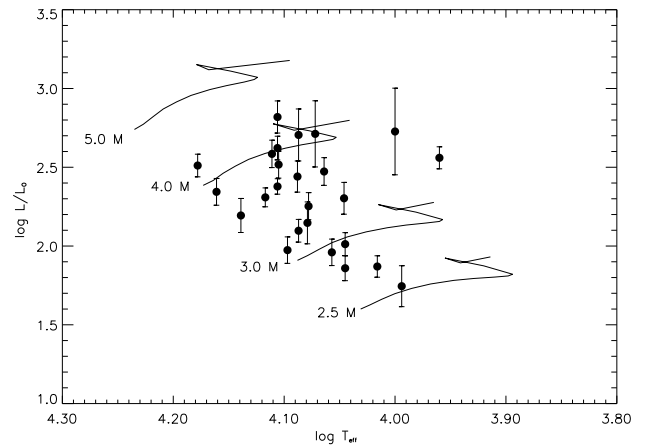}}
\caption{Position of primaries in the H-R diagram. The evolutionary tracks
computed for stars of 2, 2.5, 3, and 4 M/M$_\odot$ by Schaller et al. 
(\cite {SSMM92}) are also shown.}
\label{fig1}
\end{figure}
Since the late B-type primary always dominates in the combined
spectrum, we can safely assume that the measured spectral type is
always that of the primary. Some uncertainty in the vertical
positioning in the H-R diagram might arise by some overestimate of
the primary's luminosity due to the presence of the companion(s). In
the case of a close binary consisting of two nearly identical stars
the stellar luminosity can be overestimated by $\delta \log {\rm L}=0.3$.
However, when only faint late-type companions are 
found, the actual error is negligible, as the companion luminosity 
is fainter by more than two orders of magnitude than the primary's. 
There is still a possibility that the binary nature of the star 
induces a measurement error for the 
parallax. To test Hipparcos parallaxes on multiple stars,
Shatskii \& Tokovinin (\cite{ST98}) compared the dynamical parallaxes of 
visual
binaries with their Hipparcos trigonometrical parallaxes. No systematic
difference was found between them. The parallaxes have been found in good
agreement for all distant ($\pi<15$ mas) systems, while at intermediate 
distances ($15<\pi<30$ mas) authors found cases of large errors in Hipparcos
parallaxes attributable to short-period binaries that have been disregarded.
Of all systems in our sample only three systems are distant systems. 
The closest system with a late B-type binary has a parallax $\pi$ = 22.55 mas.

\subsection{Fundamental parameters of companions}
{\footnotesize

\begin{table*}
\caption{Fundamental parameters for companions using
evolutionary models from Baraffe et
al. (\cite{BCA98})}
\label{tab5}
\begin{tabular}{|c|c|c|c|c|c|c|c|c|c|c|}
\hline
\multicolumn{11}{|c|}{Detected visual systems}\\
\hline

HD & Comp. & M$_J$ & M$_H$ & M$_K$ & M/M$_\odot$ & q &$\rho$(AU)& $\log(L/L_\odot)_B$
  & $\log(T_{\rm eff})_B$ & $\log$(L$_X$/L$_{\rm bol})_B$\\
    \hline
\hline
HD 1685& B & 6.8 & 5.4 & 5.2 &0.60&0.22 & 214.1 &$-1.20$ &3.590&$-2.60$ \\
\hline
HD 21790$^{\mathrm{a}}$ & B & & &10.8 &0.04& 0.01& 1647.8& $<-3.83$ & 3.297&   0.07  \\
\hline
HD 27376 & B & 7.0 & 6.3 & 6.2 & 0.45 & 0.12& 291.0 & $-1.46$& 3.554 & $-3.42$  \\
\hline
HD 29589$^{\mathrm{a, b}}$ & B & & & 12.2 & & &1057.0&&& \\
\hline
HD 32964$^{\mathrm{c}}$ & B & 5.36 & 4.70 &4.71 & 0.67 & 0.23 &52.6&&&\\
\hline
HD 33802 & B&2.58 &2.56 &2.88 &&&27.4&&&\\
\hline
HD 73340 & B & 3.54 &2.93 & 2.87 & 1.20 & 0.36 &86.4&$-0.15$ & 3.647 & $-3.09$  \\
\hline
HD 73952 & B & 5.85 & 5.30 & 4.87&0.62& 0.19&179.9& $-1.06$ & 3.609& $-3.02$ \\
\hline
HD 75333 & B & 4.57& 3.78 & 3.81&0.88& 0.24&179.8& $-0.30$ & 3.695& $-3.00$ \\
\hline
HD 110073& B & 3.11 & 2.82 & 2.75 &1.13 &0.28&129.7& 0.09 & 3.753 & $-4.10$ \\
\hline
HD 114911 & B & 4.84 & 4.14 & 3.96&0.88& 0.22&336.6& $-0.49$ & 3.690 & $-3.14$\\
\hline
HD 133880 & B & 3.50& 3.04 &2.90 &1.17 &0.37 &154.7&0.15 & 3.761 & $-4.38$\\
\hline
HD 134837$^{\mathrm{a}}$ & B && &5.76&0.18& 0.06&521.7&$-1.55$ & 3.508 & $-2.68$\\  
\hline
HD 134946$^{\mathrm{a}}$ & B & & &7.00&0.30& 0.08&1038.0& $-2.18$ & 3.432 & $-1.62$\\  
\hline
HD 135734$^{\mathrm{a, c}}$  & B && &0.57 &&&92.5&&&  \\
           & C && & 9.95 & 0.05&0.01&547.6&$-3.37$ & 3.368 & $-0.10$\\
\hline
HD 145483 $^{\mathrm{a, c}}$&  B & & &2.05&&&344.3&&&\\
          &  C && &3.14& 1.08& 0.37& 18.5&$-0.02$  &3.741 & $-3.26$\\
\hline
HD 159376$^{\mathrm{a}}$ & B  & &  &5.50& 0.54& 0.13&2358.5& $-1.35$ & 3.573& $-1.74$\\
          & C && &7.02&0.28& 0.07& 1468.8&$-2.01$ & 3.531& $-1.09$\\
          & D && &7.91&0.16& 0.04& 1718.1& $-2.40$ & 3.505& $-0.69$ \\ 
\hline
HD 169022$^{\mathrm{c}}$ & B &3.44& 3.22 & 3.27&&&106.0& & & \\
\hline
HD 169978$^{\mathrm{a}}$&B & &  &7.86& 0.15 &0.03&452.9& $-2.40$ & 3.504 & $-1.20$\\
\hline
\hline
\multicolumn{11}{|c|}{Stars already known to be members of visual binary 
systems}\\
\hline
HD 27657$^{\mathrm{a, c}}$& B & & & 2.4&&&586.6&&&\\  
\hline
HD 36408$^{\mathrm{a, d}}$& B & & & -1.3&&&3336.6&&&\\  
\hline
HD 78556$^{\mathrm{a, c}}$ & B &&&1.76&&& 358.2&&& \\
\hline
HD 90972$^{\mathrm{a, c}}$ & B &&&2.50&&&1634.9&&&\\     
\hline
HD 119055$^{\mathrm{a}}$ & B &&&3.37&1.02&0.40&435.0& $-0.12$ & 3.730& $-3.53$ \\
\hline
HD 184707$^{\mathrm{a}}$ & B &&&3.77&0.92&0.33& 141.2& $-0.36$ & 3.700& $-3.01$\\
\hline
\end{tabular}         
\begin{list}{}{}
\item[$^{\mathrm{a}}$] We only have K-band photometry. Thus, the 
absolute J and H-magnitudes entries are left open.
\item[$^{\mathrm{b}}$] The companion has a very low absolute K-magnitude and 
cannot be fitted by evolutionary models.
\item[$^{\mathrm{c}}$] Values for fundamental parameters are presented in 
Tab.~\ref{tab6}.
\item[$^{\mathrm{d}}$] The magnitude difference between the primary and the
companion in the K-band is only 0.25 mag. Gravitationally bound companion
would be a main sequence star with the mass similar to that of primary.
\end{list}
\end{table*}}

{\footnotesize

\begin{table*}
\caption{Fundamental parameters for the companions on the main sequence}
\label{tab6}
\begin{tabular}{|c|c|c|c|c|c|c|c|}
\hline
\multicolumn{8}{|c|}{Detected visual systems}\\
\hline
\hline

HD & Comp. & M/M$_\odot$ & q & $\log(L/L_\odot)$ & $\log(T_{\rm eff})$ &
    $\log(L_X/L_{\rm bol})$ & Stellar model used\\
\hline
\hline
HD 32964 & B &$ 0.67$& $ 0.23$& $ -0.94$ & $ 3.640$ & $\leq -2.88$& Charbonnel et al. (\cite{CDS99})\\
\hline
HD 33802 & B &$\geq 1.05$& $\leq 0.29$& $\geq -0.06$ & $\geq 3.763$ & $\leq -2.85$& Schaller et al. (\cite{SSMM92})\\
\hline
HD 135734  & B &$>1.00$& $<0.26$&$>-1.16$&$>3.751$&$<-3.31$ &Schaller et al.
(\cite{SSMM92})\\
\hline
HD 145483 &  B &$\geq$1.28 &$\leq$ 0.44& $\geq 0.37$ & $\geq 3.812$& $\leq -3.66$& Schaller et al. (\cite{SSMM92})\\
\hline
HD 169022  & B & 0.95 &0.27 &$-0.05$&3.764 &$-4.20$ & Schaller et al. (\cite{SSMM92})\\ 
\hline
\hline
\multicolumn{8}{|c|}{Stars already known to be members of visual binary 
systems}\\
\hline
HD 27657 & B &$\geq$1.18&$\geq$0.35 & $\geq 0.20$ & $\geq 3.792$ & $\leq -4.20$& Schaller et al. (\cite{SSMM92})\\  
\hline
HD 78556 & B&$\geq$1.40& $\geq 0.36$&$\geq 0.55$& $\geq 3.835$ & $\leq -3.84$ &Schaller et al. (\cite{SSMM92}) \\
\hline
HD 90972 & B &$\geq 1.15$&$\geq 0.34$& $\geq 0.14$ & $\geq 3.785$ & $\leq -3.86$& Schaller et al. (\cite{SSMM92})\\ 
\hline
\end{tabular}
\end{table*}}

In order to determine the
evolutionary state of detected companions we need to know
their position in the H-R diagram or in the M$_K$-(M$_J$-M$_K$) 
color-magnitude diagram (CMD). 

Since late B-type stars are rather young, less than a few hundred million
years old (Table 4), we expect that a significant fraction of the companions
with low masses are PMS stars or very young main sequence stars.
The masses, luminosities and effective temperatures for PMS companions
can be then obtained from the location of the companions in
M$_K$-(M$_J$-M$_K$) color-magnitude diagram using
evolutionary models for low-mass PMS stars calculated by Baraffe et
al. (\cite{BCA98}) (hereafter B98). These models include the most recent
interior physics and the latest generation of non-grey atmosphere
models. In fact, two sets of models of Baraffe et al. (\cite{BCA98}),
namely B98$_{\mathrm{LMS}}$ and B98$_{\mathrm{Sun}}$, can be used to derive
the fundamental parameters of PMS
companions. The set B98$_{\mathrm{LMS}}$ uses a helium abundance Y=0.275
for [M/H]=0 and a general mixing length parameter $\alpha$ = $l$/H$_P$
= 1.0.  It is calculated for the masses in the range from 0.020
M$_\odot$ to 1.2 M$_\odot$. The models B98$_{\mathrm{Sun}}$ were computed
to reproduce the properties of the Sun at 4.61 Gyrs, with $\alpha$ =
1.9 and Y = 0.282. It is difficult to decide which mixing
length parameter is the better one for PMS models. There are no
theoretical arguments and the few observations available so far do not
give any clear picture either.  Since the models B98$_{\mathrm{Sun}}$ are
computed only for 0.7 to 1.2 M$_\odot$ we decided to use for this work
the set B98$_{\mathrm{LMS}}$ with the wider mass range, although for some
stars the comparison was made with the B98$_{\mathrm{Sun}}$ too.  

For 11 of our systems, we have observations in three bands, J, H and
K. In Figures 2 and 3 we show the position of PMS companions in CMD for the
set B98$_{\mathrm{LMS}}$ and set B98$_{\mathrm{Sun}}$, respectively. The
comparison of both figures reveals rather conspicuous differences in
the position of the companions and, therefore, in the inferred
fundamental parameters. In four systems, \object{HD~1685}, \object{HD~32964},
\object{HD~33802} and \object{HD~169022}, the companions cannot be fitted
by PMS evolutionary tracks. It is very likely that the system \object{HD~1685}
does not form a physical pair. In the systems \object{HD~32964},
\object{HD~33802} and \object{HD~169022}, the new companions discovered are in
an advanced evolutionary
state, i.e. they have already evolved to the main sequence and cannot be fitted
by PMS evolutionary tracks.  Their position on the main sequence can
be obtained by using evolutionary models for main-sequence stars
calculated by Schaller et al. (\cite{SSMM92}, hereforth Sch92) or by
Charbonnel et al. (\cite{CDS99}) for masses below 0.8~M$_\odot$. However, to 
apply these models to our
companions we need their magnitudes in the visual bandpass, which are unknown.
Therefore, we first estimated  the masses of the
companions in an empirical way using M$_{\rm (JHK)}$-mass relations
(MLR) of Henry \& McCarthy (\cite{HC93}).  These relations were
determined for stars with masses 0.08 to 1.0 M$_\odot$.  Under the
assumption of coeval formation of studied systems, the goal was to
place in the H-R diagram the companions with an empirical mass estimated from
the M$_{\rm (JHK)}$ MLR of Henry \& McCarthy (\cite{HC93}), along the
isochrones of the B primaries. In this way we were able to estimate their
masses, luminosities and effective temperatures.

For 14 systems the only available information are
our K-band data. Knowing the age of late B-type primaries and assuming
that systems are coeval, we placed the companions with known M$_K$
values along the isochrones of the B primaries to estimate their
masses, luminosities and effective temperatures.

\begin{figure}
\resizebox{\hsize}{!}{\includegraphics[width=9.3cm]{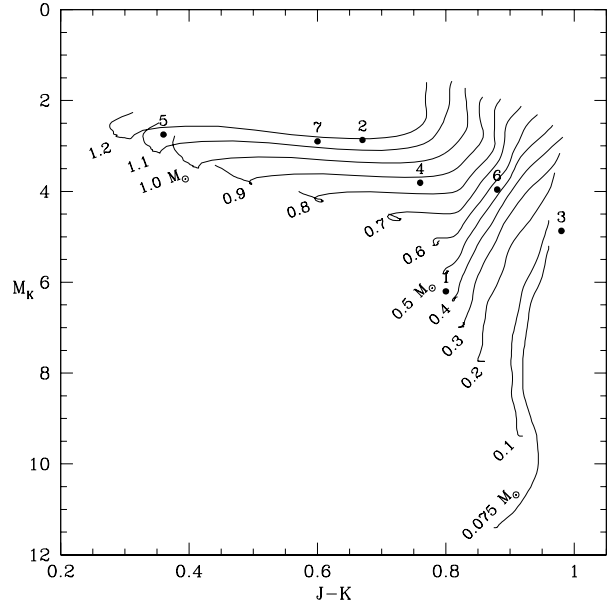}}
\caption{Position of companions in the M$_K$-(M$_J$-M$_K$)
color-magnitude diagram for PMS stars using the set B98$_{\mathrm{LMS}}$.
Indicated stars: HD~27376[1];
HD~73340[2]; HD~73952[3]; HD~75333[4]; HD~110073[5]; HD~114911[6]; 
HD~133880[7].}
\label{fig2}
\end{figure}

\begin{figure}
\resizebox{\hsize}{!}{\includegraphics[width=9.3cm]{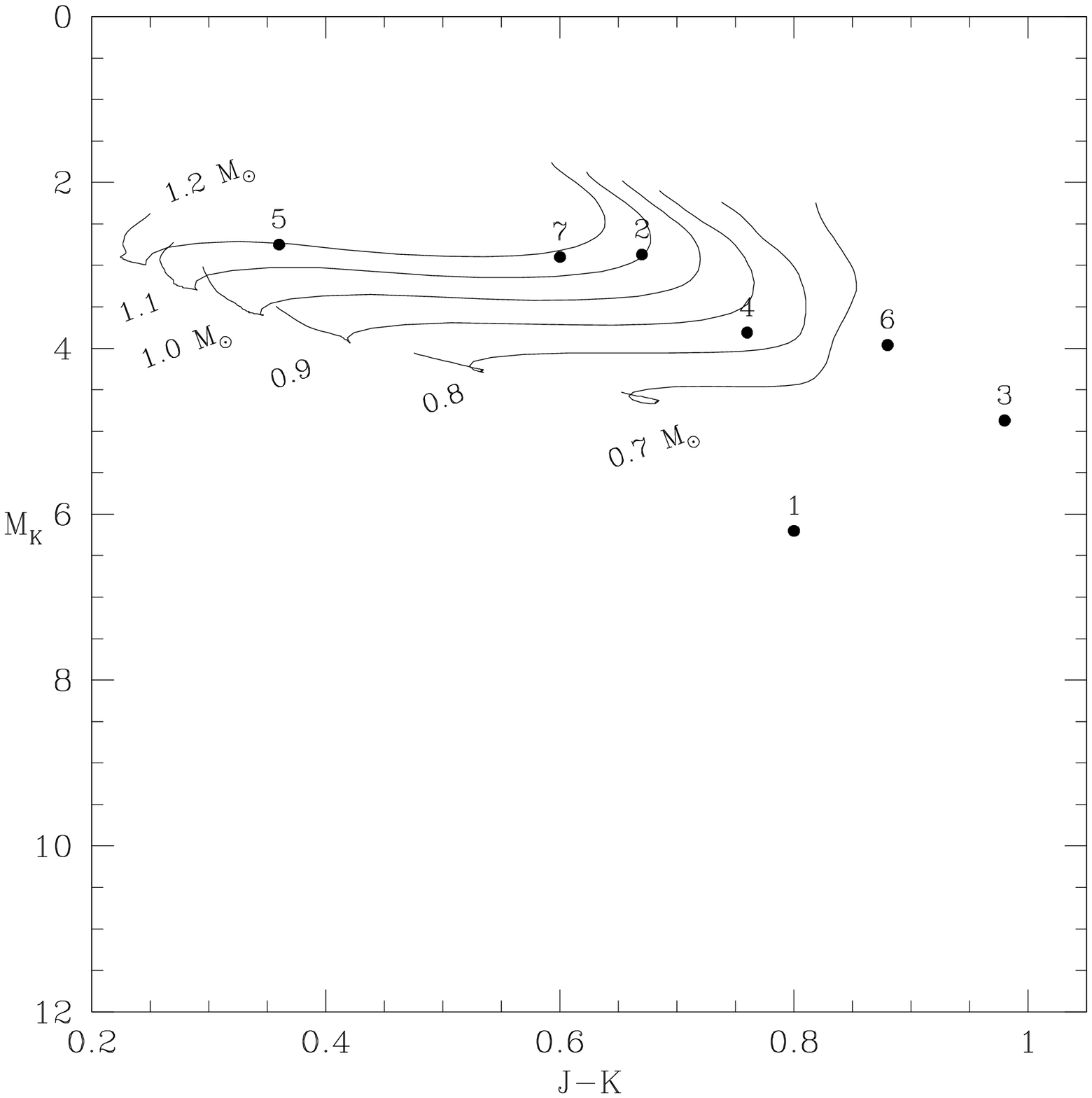}}
\caption{Position of companions in the M$_K$-(M$_J$-M$_K$)
color-magnitude diagram for PMS stars using the set B98$_{\mathrm{Sun}}$.}
\label{fig3}
\end{figure}

Further constraints on the companion properties can be obtained
from the saturation
limits of $\log({\rm L}_X/{\rm L}_{\rm bol})$ found for late-type 
coronal sources. 
They can emit up to $\approx 10^{-3}$ of their total luminosity in the soft 
X-ray band (Schmitt \cite{Sch97}). For example, a study of X-ray-discovered
T Tauri stars in the
Taurus-Auriga T Association (Neuh\"auser et al. \cite{NSS95}) revealed
${\rm L}_X/{\rm L}_{\rm Bol}$ values of $\log({\rm L}_X/{\rm L}_{\rm
bol}) = -4.06$ for G-type and $-3.56$ for M-type TTS.
The studies of nearby young open clusters show a considerable spread in coronal 
activity for stars with different rotational velocities. Although most
stars appear to saturate at about $10^{-3}$ of the stellar bolometric 
luminosity, a few stars reach higher X-ray activity levels, up to 
$\log({\rm L}_X/{\rm L}_{\rm bol}) \approx $-$2.9$\dots$ $-$2.8$~
(Stauffer et al.~\cite{SHP97}; Stauffer et al.~\cite{SCG94};
Randich et al.~\cite{RSP95}). To place a lower limit on
the mass of the companion, we use the condition that the ratio
$\log({\rm L}_X/{\rm L}_{\rm bol})$ does not exceed -3.

In cols. 2, 3 and 4 of Table 5 we present absolute J, H and K magnitudes for 
the companions. 
The masses of PMS companions and
corresponding luminosities and effective temperatures were obtained by
interpolation in the evolutionary tracks of the models
B98$_{\mathrm{LMS}}$. 
As our systems are located at rather short distances, they are unlikely to 
exhibit significant interstellar reddening.
Nevertheless, we have computed the colour excess $E[B-V]$ of the B stars in
the Geneva system using the calibration of Cramer (\cite{C82}) and obtained the
more widely used $E(B-V)$ colour excess of Johnson's system through the relation
$E(B-V)=0.842\,E[B-V]$ (Cramer \cite{C84}). One object not measured in the
Geneva system but having $uvby\beta$ colours in the literature
(\object{HD 119055}) has
$E(b-y)=0.028$ according to Crawford's (\cite{C78}) calibration, which implies
$E(B-V)=0.036$. In this way we could verify that all
but 4 stars have $E(B-V)\leq 0.05$, the others having $E(B-V)\sim 0.08$ to
$0.17$. The largest colour excess is that of \object{HD 159376}, a Si star,
so it is not very reliable because of the anomalous colours. In this worst case,
the extinctions in the $J$, $H$ and $K$ bands are respectively 0.15, 0.09 and
0.06~mag.
Therefore null extinction was 
assumed in calculation of absolute  JHK magnitudes. 
In column 6 we give the companion masses interpolated from theoretical
evolutionary models calculated by Baraffe et al. (\cite{BCA98}).
The mass-ratios in the systems are presented in column 7.
From the B98 models inferred luminosities and effective temperatures 
are given in columns 8 and 9. The last column displays the 
expected ratios $\log({\rm L}_X/{\rm L}_{\rm Bol}$).

In Table 6 we present fundamental parameters of the companions on the main 
sequence. For companions with masses $\geq$0.8 M$_\odot$ we used evolutionary
tracks of Schaller et al. (\cite{SSMM92}), which cover the range of 
0.8 to 120 M$_\odot$, to estimate 
$\log(L/L_\odot)$, $\log(T_{\rm eff})$ and $\log(L_X/L_{\rm bol})$.
For companions with lower masses we used the evolutionary tracks of Charbonnel
et al. (\cite{CDS99}).
As in the paper of Bergh\"ofer et al. (\cite{BSC96}) we adopt for calculation
of $\log(L_X/L_{\rm bol})$ a typical error of $~$0.3 dex.

\subsection{Notes on stars}
In the following we give a brief overview of the studied systems.

{\it \object{HD~1685} = HR~83}: We cannot fit the companion with the
B98 models. It is quite possible that this system does 
not form a physical pair.
Using the magnitude  M$_K$ alone and
assuming the coeval formation of the system we obtain for the mass of the
companion the value 0.6 M$_\odot$. 
We calculate a fractional X-ray luminosity of
$\log({\rm L}_X/{\rm L}_{\rm Bol}) = -2.60$ which is above the saturation limit.
{\it \object{HD~21790} = HR~1070}: This system is the widest system in our sample.
No images with the J and H filters were taken and the probability to find a
background star of such a faint magnitude of 16.2 is high.
If we assumed that this system is a physical pair,
the mass of the companion inferred from the absolute K-magnitude would be very 
low (M = 0.04 M$_\odot$) and the companion would be expected to be 
a brown dwarf. Another problem is that in either case (background object or
physical pair) 
the companion discovered here cannnot produce the observed X-ray luminosity.
We suggest that the X-radiation very likely originates from an
additional close unresolved late-type companion with coronal X-ray
emission.

{\it \object{HD~27376} = HR~1347}: This star is an SB2 star with an orbital
period of 5.0 days and a mass ratio close to $1.0$ (Batten et al. \cite{BFM89}).
The estimated angular separation of the components of this system 
is 0.\arcsec0016 (Halbwachs \cite{H81}).
According to the B98$_{\mathrm{LMS}}$ model the detected companion is a PMS 
star. Two additional companions at 49.\arcsec4 and at 0.\arcsec2 are mentioned
in the Hipparcos Input Catalogue (CDS Catalogue I/196), (henceforth HIC) and in
the BSC. Thus, this system could consist of five stars.
We calculate a fractional X-ray luminosity of
$\log({\rm L}_X/{\rm L}_{\rm Bol}) = -3. 42$. This value is well below the
saturation limit of 
$\log({\rm L}_X/{\rm L}_{\rm Bol})\approx -3$.

{\it \object{HD~29589} = HR~1484}: The companion has a very low M$_K$ magnitude
(12.2) and cannot be fitted by B98 models. No 
images with the J and H filters were taken and the probability to find a
background star of such a faint magnitude is very high.

{\it \object{HD~32964} = HR~1657}: This star is an SB2 star with an orbital
period of 5.5 days and a mass ratio close to
$1.0$ (Batten et al. \cite{BFM89}). The estimated angular separation of the 
component of this system 
is 0.\arcsec0012 (Halbwachs \cite{H81}).  
We cannot fit the detected companion with the model B98$_{\mathrm{LMS}}$.
Therefore, we assume that it is rather a main sequence star
with a mass of 0.67~M$_\odot$ computed from the MLR relation of
Henry \& McCarthy (\cite{HC93}), even though it might be a $\sim 0.7$~M$_\odot$
PMS star according to the B98$_{\mathrm{Sun}}$ model. The companion's 
fractional X-ray luminosity amounts to
$\log({\rm L}_X/{\rm L}_{\rm Bol}) = -2.88$, a value which is close to 
the saturation limit of $\log ({\rm L}_X/{\rm L}_{\rm Bol}) \approx -3$. 
An additional companion at the angular distance of
52.\arcsec2 is mentioned in the HIC and in the BSC. 
Hence, this system 
may be, in fact, a quadruple system. 

{\it \object{HD~33802} = HR~1696}: The companion is not a PMS star. Using 
the M$_{\rm (JHK)}$-mass relations given by
Henry \& McCarthy (\cite{HC93}) for main-sequence stars we estimate
the mass of the companion to be $\geq 1.05$~M$_\odot$. As these 
relations were determined only for stars with masses 0.08 to 1.0 M$_\odot$, 
only a lower limit of the mass of the companion can be estimated. As far 
as we know, the infrared MLRs have not been extended to higher masses through
other studies. 
According to the Sch92 models, the detected
companion could be a very young main-sequence star 
with a fractional X-ray luminosity of 
$\log({\rm L}_X/{\rm L}_{\rm Bol})\leq -2.85$.
An additional companion at 12.\arcsec1 is mentioned
in the HIC and in the BSC. Therefore, this system could 
be a triple system.

{\it \object{HD~73340} = HR~3413}: According to the B98$_{\mathrm{LMS}}$ models
the companion is a PMS star with a mass of 1.20 M$_\odot$. This system belongs
to the open cluster IC 2391 (BSC).

{\it \object{HD~73952} = HR~3435}: We cannot fit the companion with the
B98 models. It is quite possible that this system does 
not form a physical pair.
Using the magnitude  M$_K$ alone and
assuming coeval formation of the system we obtain for the mass of the
companion the value 0.62 M$_\odot$. 
We calculate a fractional X-ray luminosity of
$\log({\rm L}_X/{\rm L}_{\rm Bol}) = -3.02$
which is very close to the saturation limit. The star \object{HD 73952}
belongs to the open cluster IC 2391 (BSC).
An additional companion at 22.\arcsec3 is mentioned in the
HIC and in the BSC.

{\it \object{HD~75333} = HR~3500}: According to the B98$_{\mathrm{LMS}}$ models
the companion is a PMS star. We calculate a fractional X-ray luminosity of 
$\log({\rm L}_X/{\rm L}_{\rm Bol}) = -3.00$, which is at the level of
the saturation limit.
This system possibly belongs to the Pleiades Group (BSC).

{\it \object{HD~110073} = HR~4817}: This star is an SB1 star (Schneider \cite{S81}). 
According to the 
B98$_{\mathrm{LMS}}$ models the companion is a PMS star. This system belongs 
to the Pleiades group (BSC).

{\it \object{HD~114911} = HR~4993}: This is an SB2 star with an orbital period
of 20.0 days and a mass ratio close to
one (Batten et al. \cite{BFM89}). 
The mass of the companion obtained by interpolation in B98$_{\mathrm{LMS}}$ 
models is 0.57  M$_\odot$. However the age of the companion with such a mass
based on isochrone fitting (0.01$\cdot 10^8$) is in strong disagreement
with the age of the primary star (1.31$\cdot 10^8$).
Using the magnitude  M$_K$ alone and
assuming the coeval formation of the system we obtain for the mass of the
companion the value 0.88 M$_\odot$. 
We calculate a fractional X-ray luminosity of 
$\log({\rm L}_X/{\rm L}_{\rm Bol}) = -3.14$
As an additional companion at 60.\arcsec2 is mentioned in the HIC and in the
BSC, this system could be a quadruple system.

{\it \object{HD~133880} = HR~5624}: According to the B98$_{\mathrm{LMS}}$ models
the companion is a PMS star with a mass of 1.17 M$_\odot$.
This system belongs to the Scorpius-Centaurus OB association (BSC).

{\it \object{HD~134837} = HR~5653}: No images  were taken 
for this star with the J and H filters. 
If we assume that the companion is gravitationally bound and 
not a background object, we obtain from B98$_{\mathrm{LMS}}$ models that the
mass of the companions is 0.18 M$_\odot$.
We calculate a fractional X-ray luminosity of 
$\log({\rm L}_X/{\rm L}_{\rm Bol}) = -2.68$ which is above 
the saturation level.

{\it \object{HD~134946} = HR~5655}:   No images  were taken 
for this star with the J and H filters. If we assume a coeval 
formation of this system we 
obtain, according to the B98$_{\mathrm{LMS}}$ models,
that the companion 
is a PMS star with a mass of 0.30 M$_\odot$. We have derived a fractional X-ray 
luminosity of 
$\log({\rm L}_X/{\rm L}_{\rm Bol}) = -1.62$. This value is much larger than 
the saturation limit of $\log ({\rm L}_X/{\rm L}_{\rm Bol}) \approx -3$.
It is possible that the X radiation might originate 
from an additional close, unresolved low-mass companion with coronal X-ray 
emission.

{\it \object{HD~135734} = HR~5683}: Only K imaging has been done for this system. 
Under the assumption of a coeval formation we obtain, according to 
the B98$_{\mathrm{LMS}}$ models, that the companion at the angular distance of 
6.\arcsec15
is a PMS star of the mass 0.05M$_\odot$.
Given the very low mass inferred from the absolute 
K-magnitude, the faint
companion could be a brown dwarf. We have derived a fractional X-ray 
luminosity of $\log({\rm L}_X/{\rm L}_{\rm Bol}) = -0.16$. This value is 
much larger than the saturation limit of 
$\log ({\rm L}_X/{\rm L}_{\rm Bol}) \approx -3$.
This system is likely a quadruple system.
The companion observed at 1.\arcsec0 was already known (HIC \& BSC) and, 
according to B98 
models, it is not a PMS star. We estimate for this companion a mass
$>1.00$ and calculate a fractional X-ray 
luminosity of
$\log({\rm L}_X/{\rm L}_{\rm Bol})$ $<-3.31$. 
An additional companion at 23.\arcsec5 is mentioned in the HIC and in the BSC.
We suggest that X-radiation very likely originates either from the
already known  companion at 1.\arcsec0 or from the companion at the angular 
distance of 23.\arcsec5.
This system belongs to the Scorpius-Centaurus OB association (BSC).

{\it \object{HD~145483} = HR~6029}: No images were taken with the J and H filters. 
The companion at the angular distance of 3.\arcsec8 was
already known (HIC \& BSC).  According to the B98$_{\mathrm{LMS}}$
models it is not a PMS star.
Using M$_{\rm (JHK)}$-mass relations given by
Henry \& McCarthy (\cite{HC93}) for main-sequence stars, we estimate
that the mass of the companion is $\geq 1.28$.
According to the Sch92 models the known companion is a  ZAMS star. 
The companion discovered at the angular distance of 0.\arcsec2 has a mass 
M = 1.08 M$_\odot$ and is a PMS star according to the B98$_{\mathrm{LMS}}$
models. This triple system belongs to the upper Scorpius region (BSC).

{\it \object{HD~159376} = HR~6545}: . Only K imaging has been done for this system.
The field around this star is very crowded 
and the probability of finding a background star is high.
If we assume the coeval formation of this system, we 
obtain, according to the B98$_{\mathrm{LMS}}$ models,
that all three companions are PMS stars. The calculated 
fractional X-ray luminosity of all three companions is much larger than 
the saturation limit of $\log ({\rm L}_X/{\rm L}_{\rm Bol}) \approx -3$.
In other words, the detected companions alone cannot produce the observed X-ray 
luminosity.

{\it \object{HD~169022} = HR~6879}: According to the B98 models the companion is not a
PMS star. From the empirical mass estimate using the M$_{\rm (JHK)}$-mass 
relations given by Henry \& McCarthy (\cite{HC93}) we obtain the value
0.95 M$_\odot$. 
Using Sch92 models we find that the companion is a 
ZAMS star. An additional companion at the angular distance of 
32''.3 is mentioned in HIC and BSC.

{\it \object{HD~169978} = HR~6916}: No images  were taken with the J and H filters.
According to the B98$_{\mathrm{LMS}}$ models the 
companion is a PMS star of 0.15 M$_\odot$. The calculated 
fractional X-ray luminosity of the companions is much larger than 
the saturation limit of $\log ({\rm L}_X/{\rm L}_{\rm Bol}) \approx -3$.

{\it \object{HD~27657} = HR~1372}: No images were taken with the J and H filters.
The presence of a companion was already known before
we observed this star (HIC \& BSC). According to the B98 models the 
companion is not a PMS star. From the Henry \& McCarthy's (\cite{HC93}) models 
we obtain a lower limit for the mass M$\geq$1.18 M$_\odot$.
According to the  Sch92 models the 
companion is a very young main-sequence star. 

{\it \object{HD~36408} = HR~1847}: This is an SB star (BSC).
No images  were taken with the J and H filters. The visual 
companion was already known before we
observed this star. The magnitude difference $\Delta$K between the primary
and the companion in the K-band is very small (0.25). 
Considering the position of the primary in the H-R diagram, we conclude that
a gravitationally bound companion should be a main sequence star.

{\it \object{HD~78556} = HR~3630}: This is an SB star (BSC). 
No images  were taken in J and H. 
The visual companion was already known before our 
observations (HIC \& BSC). According to the B98 models the 
companion is not a PMS star. Models of Henry \& McCarthy give 
a lower limit of M$\geq$1.40 M$_\odot$.
According to the Sch92 model the companion could be a ZAMS star. 

{\it \object{HD~90972} = HR~4118}: This is an SB2 star (BSC) and the companion was 
known before we observed this system (HIC \&BSC).
No images in J and H were taken. 
According to the B98 models the
companion is not a PMS star. 
For the mass we get M$\geq$1.15 M$_\odot$. From the Sch92 models,
we conclude that the companion could be a ZAMS star. 

{\it \object{HD~119055} = HR~5144}: No images were taken with the J and H filters. 
The companion was already known (HIC \& BSC). According to the 
B98$_{\mathrm{LMS}}$ models 
the companion is a PMS star of the mass 1.02 M$_\odot$. 

{\it \object{HD~184707} = HR~7440}: Only K imaging has been done for this system. 
The companion was already known (HIC \& BSC). According to the 
B98$_{\mathrm{LMS}}$ models the companion is a PMS star of the mass 0.92 M$_\odot$.

\section{Discussion}\label{sec4}
The results of our study clearly confirm our suspicion that most late B-type
stars which are detected in X-rays are accompanied by a low-mass late-type
companion. Although we have no formal evidence that the X-ray emission is due
the low-mass companion (e.g. spectra showing that the companions are indeed
active, or lack of active low-mass companions in a control sample of B stars
which are not X-ray sources), this hypothesis appears strongly backed by our
data. If we admit that the late type companions are gravitationally bound in the 
studied systems, we expect that out of 19 detected companions, 15 are PMS 
stars.
Among the six already known visual binary systems, two contain a PMS companion.
A very interesting project would be to investigate these systems more 
closely with near infrared spectrographs used together with adaptive
optics systems.
Such spectroscopic studies would provide a much more accurate estimate of the 
stellar 
masses of the components through their spectral type and 
would allow to conclusively
determine whether the companion is, indeed, a PMS star or  
a background star.

The issue whether all studied systems form physical pairs or not deserves
further investigation.
The detected companions have projected separations ranging from 0.\arcsec2 to
14.\arcsec1 (18-2358 AU). We cannot rule out that the widest systems are not
physical ones. Estimates of cut-off separation for the gravitational binding
of the pairs by several authors (Chandrasekhar \cite{B44}, Bahcall et al. 
\cite{BRJ86}, Duquennoy \& Mayor \cite{DM91})
place it between $2\cdot 10^3$ and $2\cdot 10^4$ AU.
The radial velocity measurements and  proper motion studies
would tell us whether the stars in these systems are gravitationally 
bound or not. 
Out of the 49 late B-type stars in our sample, 25 have additional
companions in the field of view. Of the studied systems, 6 were already known
as visual binaries, and 19 are newly detectied by us.
The sample of the systems with companions probably consists
of 9 binary systems, 5 triple, 4 quadruple systems and one system consisting of
5 stars.
That yields an observed binary frequency of 51 \% for our star sample.

As mentioned in the Introduction, the formation mechanism of massive stars is
not well understood yet, and it is important to compare the multiplicity of
higher mass stars to that of lower mass stars. However, massive stars are less 
frequent than low mass stars and the large brightness of the massive
primaries prevents the detection of close visual low mass companions. 
Previous studies of the spectroscopic binary frequency of B stars have shown 
that on average the percentage of spectroscopic binaries is higher than among
solar type stars (e.g., Abt et al.~\cite{A90}; Morell \& Levato~\cite{ML91}).
McAlister et al. (\cite{M93}) carried out speckle interferometric observations
for 2088 OB stars and found a binary frequency of B stars of about 14\%.
Bouvier \& Corporon (\cite{BC2001}) observed a sample of 63 Herbig Ae/Be stars
among which 22 were B0--B8 stars and 34 B9--A9 stars. They found an observed
binary frequency for the two groups of stars of, respectively, 36 \% and 
42 \%. These values are slightly lower than that in the present work.

In our study, the random pairing, that we estimated to 2 cases over 49 studied
stars, might affect the observed binary frequency. On the other hand, very
recent results obtained from higher angular resolution imaging on the same 
star sample using Keck Observatory AO facilities already revealed two new
companions (among 8 targets) with a separation of the order of 50
milli-arcseconds (Le Mignant \cite{L2001}). A very likely reason for the 
higher binary frequency in the present work is that the sample has been
X--ray selected. As a consequence, the multiplicity study in our 
late-B star sample is biased towards low-mass
companions (which have strong X-ray emission).
For statistical purposes it will be of crucial importance
to study in the future a sample of late-B type stars not detected in the 
ROSAT all-sky survey.
Such a study will represent a systematic multiplicity study of
early type stars which would set important constraints 
to any theory of binary and multiple star formation.

Nearly 20\% of the late-B stars found in the ROSAT
survey belong to young stellar groups: Sco-Cen association, Sco
OB2, and the Pleiades cluster and supercluster. The remaining
X-ray emitting B stars are field stars. A further study of X-ray emission
of the systems with late B-type primaries, e.g. with the Chandra X-ray 
Observatory which offers rather high spatial resolution,
will provide a good opportunity to examine the link 
between X-ray emission and age in the stars having
different masses and different formation histories.

\begin{acknowledgements}
DLM wishes to acknowledge the support from 
the European Southern Observatory during the data reduction in
Garching.
We would like to thank Dr. Isabelle Baraffe (Ecole Polytechnique of Lyon) 
for providing her stellar evolutionary models and Dr. Fiorella Castelli for 
calculation of effective temperatures of a few late B-type stars using 
Str\"omgren photometry. Observing support during direct and
coronagraphic imaging observations from the the 3.6 m telescope team is also
gratefully acknowledged, especially Victor Merino, Dr. Franck Marchis and Dr. 
Olivier Marco (ESO La Silla/Santiago). PN acknowledges the support of the
Swiss National Science Foundation.
Finally, we wish to thank Patricia Goud\'e from Keck Observatory for the
technical editing work she did on the paper, and the referee, Dr. F. Walter,
for constructive comments.

\end{acknowledgements}

\end{document}